\documentclass[aps,pre,reprint,superscriptaddress,floatfix,longbibliography]{revtex4-2}

\usepackage[T1]{fontenc}
\usepackage[utf8]{inputenc}
\usepackage{lmodern}
\usepackage{amsmath,amssymb,bm}
\usepackage{graphicx}
\usepackage{booktabs}
\usepackage[percent]{overpic}
\usepackage[caption=false]{subfig}
\usepackage{microtype}
\usepackage{xcolor}
\usepackage[colorlinks=true,linkcolor=blue,citecolor=blue,urlcolor=blue]{hyperref}

\begin{document}

\title{Representation-Dependent Machine Learning of the Isotropic–Nematic Transition in the Lebwohl–Lasher Model}
\author{Maninder Kaur}
\email{Maninder.Kaur@uga.edu}
\affiliation{Center for Simulational Physics, The University of Georgia, Athens, Georgia 30602, USA}
\affiliation{Department of  Physics, The University of Georgia, Athens, Georgia 30602, USA}

\author{Aojie Xue}
\email{Aojie.Xue@uga.edu}
\affiliation{Center for Simulational Physics, The University of Georgia, Athens, Georgia 30602, USA}
\affiliation{Department of  Physics, The University of Georgia, Athens, Georgia 30602, USA}
\author{David P. Landau}
\email{dlandau@uga.edu}
\affiliation{Center for Simulational Physics, The University of Georgia, Athens, Georgia 30602, USA}
\affiliation{Department of  Physics, The University of Georgia, Athens, Georgia 30602, USA}
\date{\today}

\begin{abstract}
Machine-learning detection of phase transitions depends not only on the
learning algorithm, but also on whether the input representation preserves the
symmetries of the system. We examine this for the
weak first-order isotropic--nematic transition of the three-dimensional
Lebwohl--Lasher model, whose apolar and continuously degenerate nematic phase
makes raw molecular configurations a challenging input for unsupervised learning.
Principal component analysis (PCA) and a convolutional autoencoder (CAE) fail to identify
the transition from raw configurations because rotationally equivalent nematic
states can appear far apart in the input space. When the same configurations
are transformed into a rotationally invariant local-correlation
representation, both methods recover transition-sensitive signatures and bimodal coexistence distributions. 
A supervised three-dimensional convolutional neural network (CNN), by
contrast, accurately predicts the scalar order parameter from raw
configurations when given order-parameter labels. The Lebwohl–Lasher model therefore separates unsupervised phase discovery from supervised order-parameter regression and shows that symmetry-respecting input representations are needed for unsupervised machine learning in orientationally ordered systems.

\end{abstract}

\maketitle
\section{Introduction}
\label{sec:introduction}
 
Machine learning (ML) has emerged as a powerful complement to traditional
computational methods in condensed matter physics and statistical physics over the last decade, especially for the identification of phases of matter and phase transitions 
~\cite{Wang2016,CarrasquillaMelko2017,vanNieuwenburg2017,Hu2017,Wetzel2017,maninder2026}. Supervised neural networks can classify phases directly from microscopic configurations, while unsupervised approaches can reveal low-dimensional latent variables and clustering structures without explicit labels \cite{Wang2016,CarrasquillaMelko2017,Wetzel2017,NgYang2023AutoencodersPhaseTransitions,maninder2026}. These methods have been applied to a wide range of lattice and off-lattice systems, including Ising, Potts, XY, fermionic, and colloidal models \cite{Hu2017,Jadrich2018a,Jadrich2018b,Broecker2017,Shiina2020}. 
These studies have established ML as a probe of phase structure, but they also raise the
question of when generic learning algorithms identify physics-aware collective variables and when they instead organize data according to representation-dependent similarities.

In classical spin models, one route has been to replace raw spin configurations by correlation-based representations that account for symmetry-related configurations in systems such as Potts and clock models~\cite{Shiina2020}. Liquid crystals are a non-trivial setting for this question: their ordered phases are described by orientational order parameters and apolar continuous symmetries rather than by scalar spin variables. ML work on liquid-crystal systems has so far focused on optical textures and image-based prediction of phases or material properties. Convolutional neural networks have been used to predict liquid-crystal phases and order parameters from optical images, and recent reviews summarize the increasing use of ML in liquid-crystal phases, textures, defects, and physical-property prediction~\cite{Sigaki2019PRE,Sigaki2020,Piven2024}.

The Lebwohl--Lasher (LL) model is one of the classic lattice models for nematic liquid crystals~\cite{LebwohlLasher1972}. In three dimensions, it exhibits an isotropic--nematic transition that is weakly first order. Recent high-precision Monte Carlo work has further clarified the finite-size scaling properties of this transition~\cite{Xue2024}.
Despite this central role, a systematic ML analysis of the three-dimensional LL transition at the level of microscopic molecular configurations has remained largely absent. 
The closest LL-related ML studies have treated simulated liquid-crystal textures rather than the underlying three-dimensional molecular configurations. Sigaki \textit{et al.} used convolutional neural networks  to classify phases and predict order parameters from optical texture images generated from nematic simulations~\cite{Sigaki2020}, and generative ML has been applied to the isotropic–nematic transition in off-lattice systems such as Gay–Berne ellipsoids~\cite{Beyerle2025}. None of these studies addresses how unsupervised and supervised methods behave when applied directly to microscopic LL configurations.

The 3D LL model is, therefore, not just another lattice benchmark for ML. It poses a representation problem that is absent in scalar spin systems. Nematic order is
apolar and continuously degenerate: reversing the molecular axis does not change the nematic state, and the global director may point in any direction. 
Configurations that represent the same nematic state can therefore look unrelated when written as raw three-component molecular-orientation fields, and the numerical distance between configurations need not reflect the distance between states in order-parameter space. This representation problem is expected to arise
more generally in orientationally ordered systems with apolar continuous
symmetry, where the microscopic orientation and its reversal represent
the same local molecular axis. Thus, the central question is not only whether ML can identify the isotropic--nematic transition, but what representation of the microscopic configurations is required for different learning methods to work reliably.

To address this problem, we compare the raw molecular-orientation field with a rotationally invariant local-correlation representation. The resulting representation removes the irrelevant global orientation while retaining the local orientational information that distinguishes isotropic and nematic states. It is consistent with the apolar symmetry of nematic molecules and does not use the global scalar order parameter or phase labels. Using the 3D LL model as a test case, we analyze statistically independent Monte Carlo configurations using three approaches: principal component analysis (PCA) as a linear unsupervised method, a three-dimensional convolutional autoencoder (CAE) as a nonlinear unsupervised method, and a supervised three-dimensional convolutional neural network (CNN) for order-parameter regression. 
For the two unsupervised methods, we use two inputs derived from the same configurations: the raw molecular-orientation field and a rotationally invariant local-correlation representation. The CNN is trained on raw molecular configurations only, since its role is to test whether order-parameter labels can guide a network toward symmetry-relevant features when the input itself is not explicitly invariant. This comparison separates two questions: whether a symmetry-respecting representation is necessary for unsupervised phase identification, and whether supervision can circumvent the representation problem.

\par
The remainder of the paper is organized as follows.
Section~\ref{sec:Model} describes the LL model, the Monte Carlo simulation methodology, and the input representations used. 
Section~\ref{sec:MLmethods} presents the three ML approaches. 
Section~\ref{sec:results} reports and discusses the results for each method.
Section~\ref{sec:conclsuion} summarizes our conclusions.

\section{Model and Simulation Details}
\label{sec:Model}

\subsection{Model}
The Lebwohl--Lasher model consists of  molecules placed on a 
simple-cubic lattice and is described by the Hamiltonian
\begin{equation}
    H = -J \sum_{\langle i,j \rangle} P_2(\mathbf{u}_i \cdot \mathbf{u}_j),
    \label{eq:hamiltonian}
\end{equation}
where
\begin{equation}
    P_2(x) = \frac{1}{2}\left(3x^2 - 1\right)
    \label{eq:legendre}
\end{equation}
is the second-order Legendre polynomial, $\mathbf{u}_i$ is a unit vector 
representing the orientation of the molecule at site $i$, and $J > 0$ 
is a coupling parameter. The sum runs over all nearest-neighbor pairs 
$\langle i,j \rangle$. Because $P_2(x)$ is even in $x$, the interaction 
is invariant under $\mathbf{u}_i \to -\mathbf{u}_i$, consistent with 
apolar nematic symmetry. The ground state exhibits orientational (nematic) 
order in which all molecules are parallel, while the system remains free 
to rotate continuously in three-dimensional space.

The degree of nematic ordering is quantified by the traceless 
second-rank ordering tensor
\begin{equation}
    Q_{\alpha\beta} = \frac{1}{N}\sum_{i=1}^{N} 
    \frac{1}{2}\left(3u_{i\alpha}u_{i\beta} - \delta_{\alpha\beta}\right),
    \label{eq:Qtensor}
\end{equation}
where $N = L^3$ is the total number of molecules and $u_{i\alpha}$ is 
the $\alpha$-component of $\mathbf{u}_i$. The scalar nematic order 
parameter $S$ is defined as the largest eigenvalue of $Q_{\alpha\beta}$, 
and its corresponding eigenvector defines the \textit{director}, the 
mean orientation axis of the system. In the disordered isotropic phase 
$S \approx 0$, while in the ordered nematic phase $S$ takes a finite 
positive value. 

\subsection{Monte Carlo Simulations}
\label{sec:simulation}
We simulate simple cubic lattices of size $L = 50$, $60$, and $80$ 
with periodic boundary conditions, setting $J/k_{\mathrm{B}} = 1$. 
Configurations are generated using the Metropolis Monte Carlo (MC) 
algorithm~\cite{metropolis1953,landau2021guide} at $81$ temperatures uniformly spaced 
in $k_{\mathrm{B}}T/J \in [1.1147,\, 1.1307]$ in steps of $0.0002$. 
All simulations start from a random (fully disordered) initial state. 
For $L = 50$, $10^5$ Monte Carlo steps (MCS) of equilibration are followed by a production 
run of $5\times10^6$ MCS, with configurations saved every 
$5\times10^4$ MCS. Here, one MCS corresponds to $L^3$ attempted molecular updates, equivalent to one attempted update per lattice site on average.
For $L = 60$, the same $10^5$ MCS of equilibration are followed by a 
production run of $2\times10^7$ MCS, with configurations saved every 
$10^5$ MCS. 
For $L = 80$, $7\times10^5$ MCS of equilibration accommodate the 
slower relaxation near the transition, followed by a production run 
of $3\times10^7$ MCS, with configurations saved every $3\times10^5$ MCS. 
This yields $100$ statistically independent configurations per 
temperature for each lattice size, for a total of $8100$ 
configurations per size.

\subsection{Input representations}
\label{subsec:inputs}
Two input representations were constructed from each Monte Carlo configuration. The first was the raw molecular-orientation field. At each lattice site \(\mathbf{r}\), the molecular axis was represented by a three-component unit vector,
\begin{equation}
\mathbf{u}(\mathbf{r})=
\bigl(
u_x(\mathbf{r}),u_y(\mathbf{r}),u_z(\mathbf{r})
\bigr).
\end{equation}
A full lattice configuration is therefore an array of shape
$3\times L\times L\times L$, where the first index corresponds to the
Cartesian component of the molecular orientation and the remaining
three indices specify the lattice site.
The second representation was a rotationally invariant local-correlation
field, constructed from nearest-neighbor molecular orientations. Since the
correlation variable is defined on nearest-neighbor bonds using $P_2$, we
also refer to this representation as the bond-$P_2$ field.
For each site $\mathbf r$ and lattice direction
$\alpha\in\{x,y,z\}$, the local correlation variable was defined as
\begin{equation}
b_\alpha(\mathbf r)=
P_2\left[
\mathbf u(\mathbf r)\cdot \mathbf u(\mathbf r+\hat{\alpha})
\right],
\end{equation}
where $P_2$ is the second Legendre polynomial defined in Eq.~(\ref{eq:legendre}).
This construction gives three scalar fields, one for each Cartesian
lattice direction. The local-correlation representation therefore has
the same array shape, $3\times L\times L\times L$, as the raw
molecular-orientation representation.
The two representations were used to separate questions of model
architecture from questions of symmetry-compatible input representation.
The raw molecular-orientation field retains the full microscopic
orientation information, whereas the local-correlation field removes
the irrelevant global orientation while preserving nearest-neighbor
orientational alignment.

\section{Machine-learning methods}
\label{sec:MLmethods}

\subsection{Principal component analysis}
\label{subsec:pca}
Principal component analysis (PCA), a linear dimensionality-reduction
method that projects high-dimensional data onto orthogonal directions
of maximal variance~\cite{Jolliffe2016}, was used to test whether the dominant variance
directions of the configuration ensemble contain information about the
isotropic--nematic transition. For each lattice size, PCA was applied independently to two representations of the same Monte Carlo configurations: the raw molecular-orientation field and the rotationally invariant
bond-$P_2$ field. Since PCA requires vector input, each
$3\times L\times L\times L$ configuration array was reshaped into a
one-dimensional feature vector of length $3L^3$ before applying PCA.

The analysis was performed using the standard \texttt{PCA} routine in
\texttt{scikit-learn}. No scalar order-parameter labels or temperature
labels were used when fitting the principal components. 
Temperature information was used only after the PCA transformation to group the leading PCA scores by temperature and compute averages and probability distributions.
For the comparison across lattice sizes in Fig.~\ref{fig:pca}, the leading PCA score was divided by $\sqrt{3L^3}$ after the PCA transformation.
This procedure allows PCA to serve as a representation-sensitive
unsupervised diagnostic: if the transition is encoded in the geometry of
the chosen input space, it should appear in the leading PCA scores. Since
the sign of a principal component is arbitrary, the sign of the first
component was fixed only for plotting consistency.

\subsection{Convolutional autoencoder}
\label{subsec:cae}
\begin{figure*}[htb]
\centering
\includegraphics[width=\textwidth]{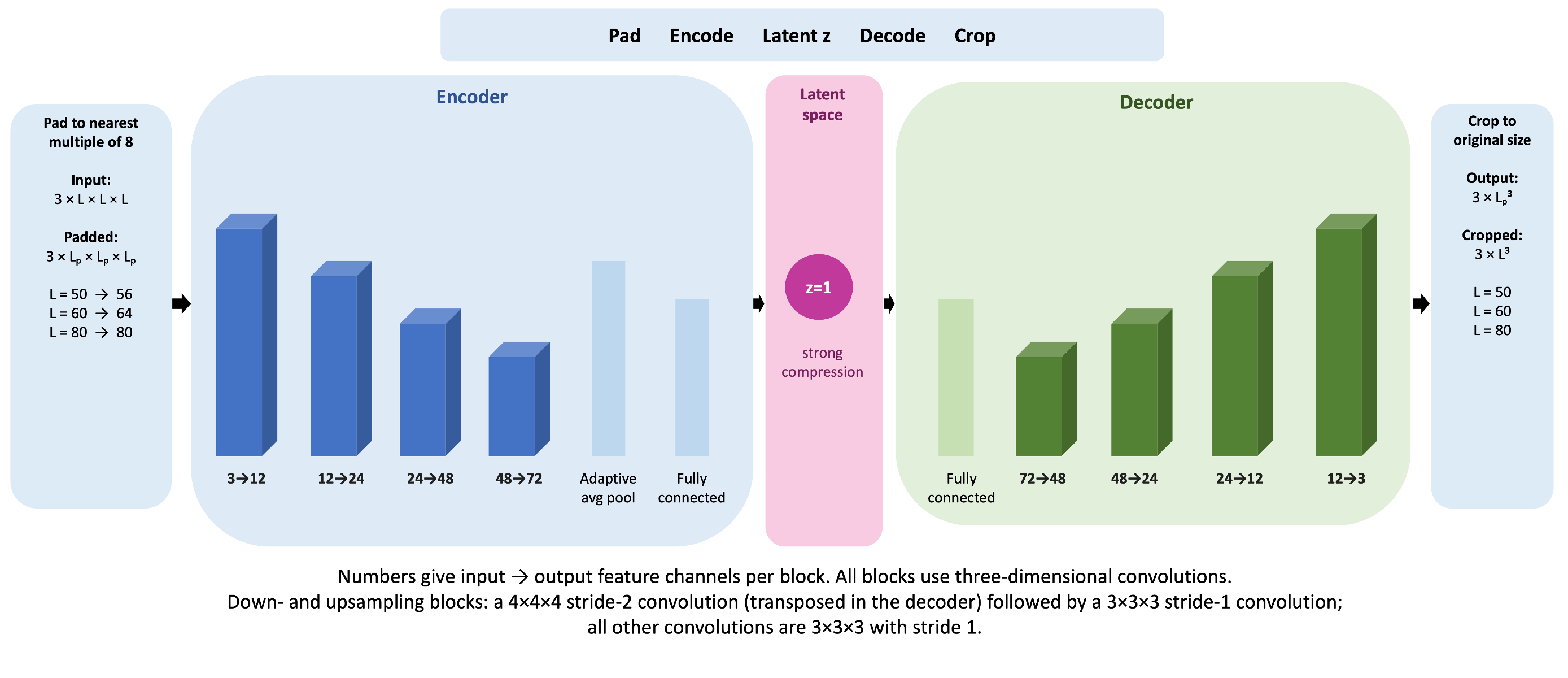} 
\caption{Architecture of the three-dimensional convolutional autoencoder (CAE) used for nonlinear unsupervised dimensionality reduction. 
}
\label{fig:CAE_arch}
\end{figure*}

An unsupervised convolutional autoencoder (CAE) is a nonlinear
reconstruction-based neural network that encodes an input into a compact
latent representation and then decodes it to reconstruct the input
\cite{Masci2011}. We used the CAE to test whether nonlinear unsupervised
learning can extract a transition-sensitive coordinate.

The CAE was applied separately to the raw molecular-orientation field
and to the rotationally invariant local-correlation field for each lattice
size. In both cases, the input was a tensor of shape
$3\times L\times L\times L$. No scalar order-parameter values were used
during training; the network was trained only to reconstruct its input.
The encoder compressed each configuration to a single latent variable
$z$. This one-dimensional bottleneck was chosen so that the learned coordinate
could be analyzed as a compact unsupervised descriptor of the configuration
ensemble. Because the CAE is optimized for reconstruction, the latent coordinate is
not assumed to be identical to the scalar nematic order parameter.
The CAE architecture is shown schematically in Fig.~\ref{fig:CAE_arch}.
Since the encoder used three stride-2
downsampling stages, each input was first padded symmetrically to
\begin{equation}
L_{\rm p}=8\left\lceil \frac{L}{8}\right\rceil ,   
\end{equation}
where  $\lceil\,\rceil$ denotes the ceiling function, so that
$L_p$ is the smallest multiple of 8 greater than or equal to
$L$. Reflective padding was used to extend the input to this
size. Thus $L_{\rm p}=56$, 64, and 80 for
$L=50$, 60, and 80, respectively.
The encoder consisted of three-dimensional convolutional blocks with batch
normalization and ReLU activations. 
Starting from the three input channels, the encoder increased the feature width through convolutional blocks with 12, 24, 48, and 72 channels.

Local feature extraction used
$3\times3\times3$ convolutions, while downsampling was performed using
$4\times4\times4$ convolutions with stride 2 and padding 1. The encoded
feature map was reduced to one value per channel using adaptive average
pooling to a $1\times1\times1$ spatial output, equivalent to global average
pooling over the final spatial dimensions. A fully connected layer then
mapped the pooled feature vector to the one-dimensional latent variable
$z$.

The decoder followed the reverse spatial progression of the encoder. A
fully connected layer first mapped the latent variable back to the encoded
tensor shape, followed by transposed three-dimensional convolutions that
restored the spatial resolution. The decoder reconstructed the padded
volume of size
$3\times L_{\rm p}\times L_{\rm p}\times L_{\rm p}$.
This output was cropped back to the original
$3\times L\times L\times L$ size before the reconstruction loss was
evaluated. The same encoder--decoder architecture was used for all lattice sizes and both input representations, with the tensor dimensions adjusted only according to $L_{\rm p}$.
The CAE was optimized by minimizing the mean-squared
reconstruction error
\begin{equation}
 {\cal L}_{\rm CAE}
=
\frac{1}{3L^3}
\sum_{c,\mathbf r}
\left[
\hat{x}_c(\mathbf r)-x_c(\mathbf r)
\right]^2 ,  
\end{equation}
where $x_c(\mathbf r)$ and $\hat{x}_c(\mathbf r)$ denote the input and
cropped reconstructed fields. 

Training was performed using the Adam optimizer~\cite{Kingma2015} with learning rate
$2\times10^{-4}$, batch size 32, and zero weight decay. No feature
standardization was applied; the raw molecular and bond-$P_2$ fields were used
in their native scales. For each lattice size and representation, the data
were split into training and validation sets using a stratified split by
temperature, with 10\% of the configurations at each temperature assigned
to validation. Temperature was used only to construct this balanced train-validation split
and was not supplied to the network as an input or target. Training was
continued for up to 50 epochs with early stopping based on the validation
loss using a patience of 20 epochs. 

After training, the latent scalar $z$ was extracted for every
configuration. The values of $z$ were then grouped by temperature to
compute temperature-resolved averages and probability distributions. As for
PCA, the overall sign of a one-dimensional latent coordinate is arbitrary;
therefore, the sign of $z$ was fixed only at the analysis stage for
consistent plotting. Since the latent dimension is one, we denote the latent coordinate simply by $z$.

For comparison across system sizes, the one-dimensional latent coordinate was
standardized separately for each lattice size as
\begin{equation}
\tilde{Z_1}=\frac{z-\langle z\rangle_L}{\sigma_L(z)},
\end{equation}
where $\langle z\rangle_L$ and $\sigma_L(z)$ were computed over all
configurations for that $L$.  This standardization was used only for
analysis and visualization, not during CAE training.

\subsection{Convolutional neural network}
\label{subsec:cnn}
\begin{figure*}[tb]
\centering

\includegraphics[width=\textwidth]{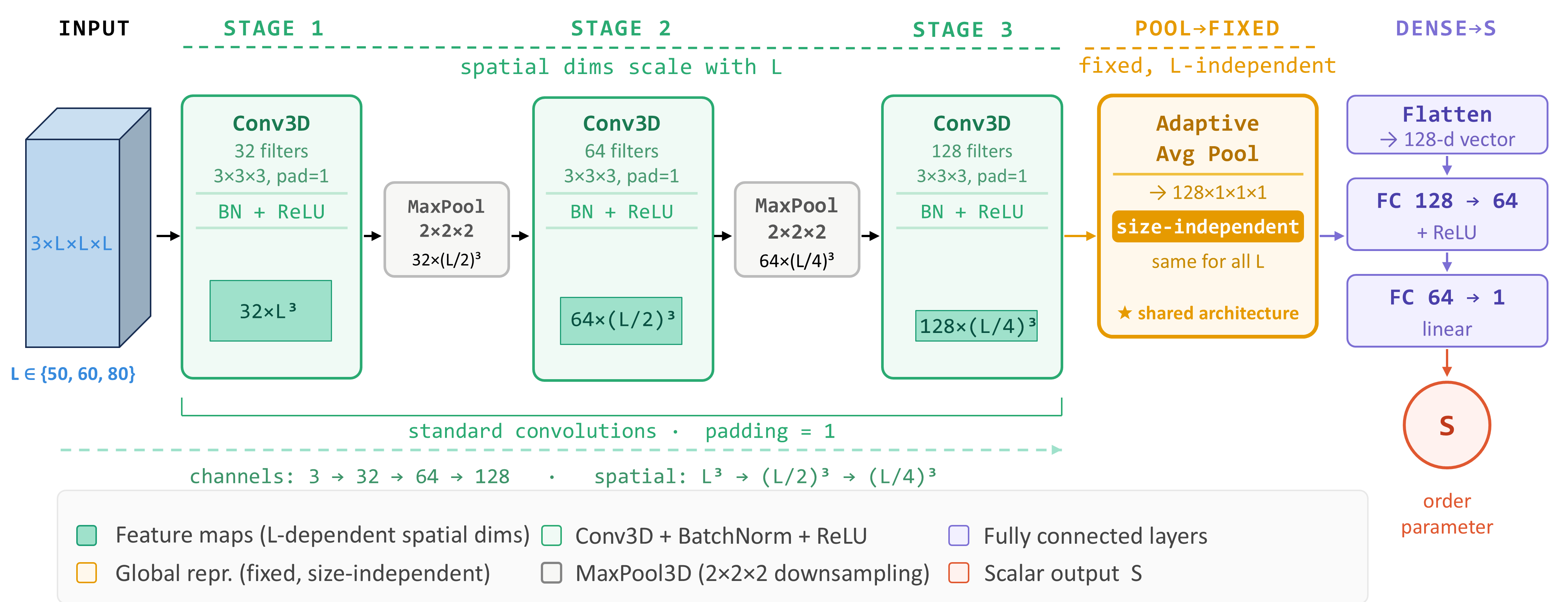} 
\caption{
Architecture of the three-dimensional convolutional neural network (CNN) used for supervised regression of the scalar nematic order parameter from raw molecular configurations.
}
\label{fig:CNN_arch}
\end{figure*}

The supervised analysis was performed using a three-dimensional
CNN trained to predict the scalar nematic
order parameter $S$ from raw Lebwohl--Lasher configurations. In contrast
to the PCA and CAE analyses, the CNN was trained only on the raw molecular
field and not on the bond-$P_2$ representation. This choice was made
because the purpose of the CNN was to test whether supervised learning
can extract symmetry-relevant information directly from raw microscopic
configurations when the scalar order parameter is supplied as a training
target.

Each input configuration was represented as a tensor of shape
$3\times L\times L\times L$, where the three channels correspond to the
Cartesian components of the molecular orientation field. The target value
for each configuration was the scalar nematic order parameter $S$ computed
independently from the ordering tensor defined in Sec.~\ref{sec:Model}A.

The CNN architecture is shown schematically in Fig.~\ref{fig:CNN_arch}.
It consisted of three three-dimensional convolutional stages followed by
a fully connected regression head. The convolutional feature extractor used
standard $3\times3\times3$ convolutions with padding 1.  The three
convolutional layers had channel widths\[3 \rightarrow 32 \rightarrow 64 \rightarrow 128 .\]
The first two convolutional stages were followed by batch normalization,
ReLU activation, and $2\times2\times2$ max pooling. The third stage was
followed by batch normalization, ReLU activation, and adaptive average
pooling to a $1\times1\times1$ output. The resulting 128-component feature
vector was passed through a regression head,
\[
128 \rightarrow 64 \rightarrow 1 ,
\]
with a ReLU activation between the two linear layers. The final scalar
output was interpreted as the predicted order parameter $S_{\rm pred}$.
For each lattice size, the data set was divided into training and test
sets using an 80/20 split stratified by temperature, so that both subsets
contained comparable coverage of the transition region. The scalar order-parameter targets were rescaled before training as
\begin{equation}
\tilde S =
\frac{S-\mu_{\rm train}}{\sigma_{\rm train}},
\label{eq:S_normalization}
\end{equation}
where $\mu_{\rm train}$ and $\sigma_{\rm train}$ are the mean and standard
deviation of $S$ over the training set. This rescaling was used only during
training to keep the regression target centered and of order unity. The
network was trained to predict $\tilde S$.
The loss function was the mean-squared error,
\begin{equation}
{\cal L}_{\rm CNN}
=
\frac{1}{N_b}
\sum_{n=1}^{N_b}
\left(
\tilde S_{{\rm pred},n}-\tilde S_n
\right)^2,
\label{eq:cnn_loss}
\end{equation}
where $N_b$ is the batch size.

For each lattice size, a separate CNN was trained using the same
architecture. Training was performed with batch size 16 using the Adam
optimizer with learning rate $10^{-4}$. After training, the predicted
normalized values were transformed back to the original scale according to
\begin{equation}
S_{\rm pred}
=
\sigma_{\rm train}\tilde S_{\rm pred}
+
\mu_{\rm train}.
\label{eq:S_inverse_transform}
\end{equation}
The regression performance was evaluated on the held-out test set using
the coefficient of determination $R^2$.

\section{Results and Discussion}
\label{sec:results}
The results are organized to follow the logic of the comparison: first the
thermodynamic benchmark, then the failure of unsupervised learning on raw
configurations, the recovery obtained from the symmetry-compatible
bond-$P_2$ representation, and finally the supervised CNN result.

\begin{figure*}[tb]
\centering
\includegraphics[width=0.9\textwidth]{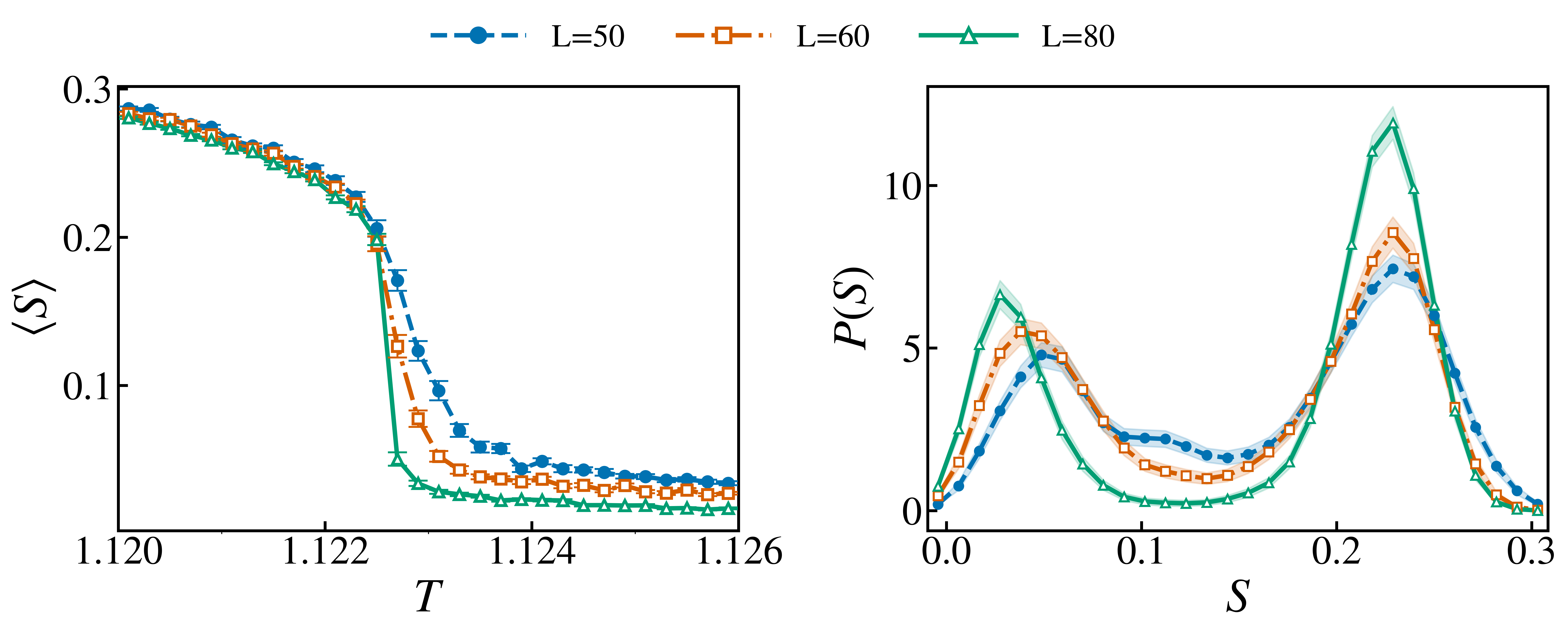} 
\caption{
Thermodynamic benchmark for the isotropic--nematic transition.
(a) Temperature dependence of the scalar nematic order parameter
$\langle S\rangle$ for $L=50$, 60, and 80. (b) Near-coexistence
probability distribution $P(S)$ constructed from the finite-size transition
windows defined in the text.}
\label{fig:thermo}
\end{figure*}

\begin{figure*}[tb]
\centering
\includegraphics[width=0.9\textwidth]{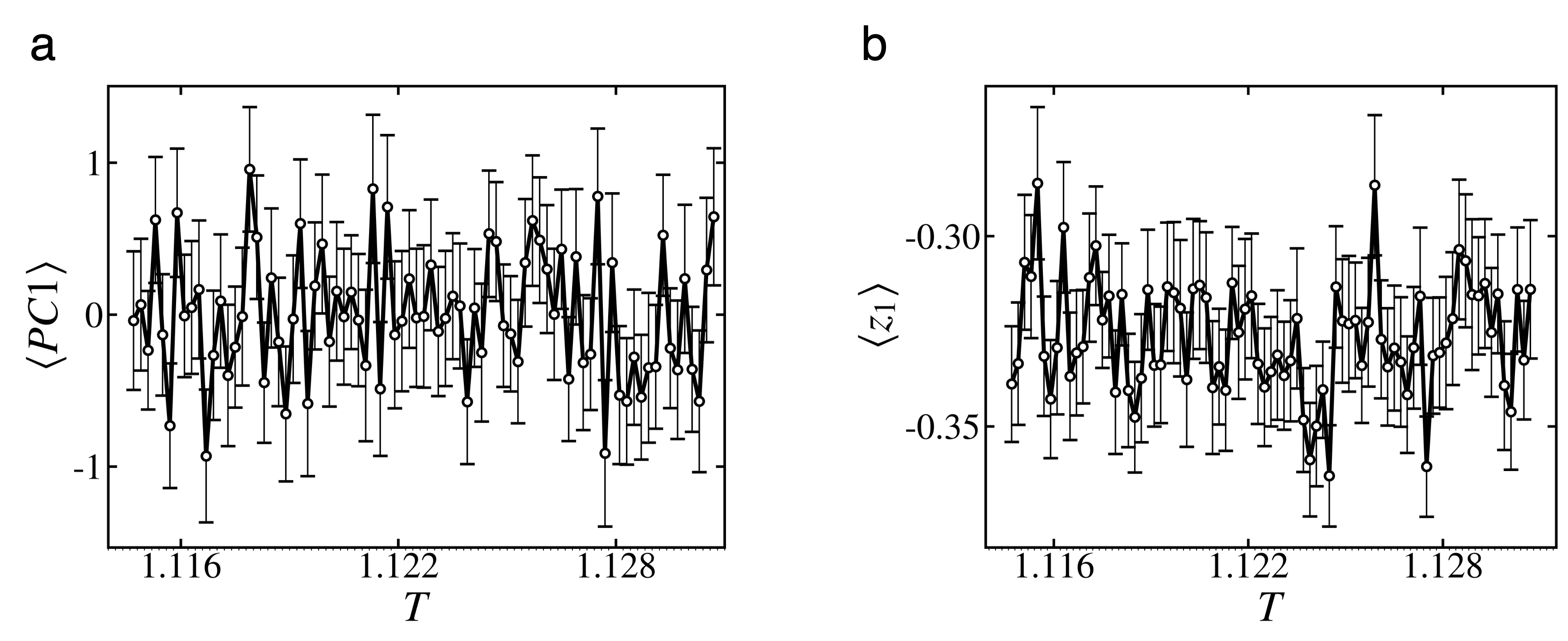} 
\caption{
Unsupervised learning on raw molecular configurations. Representative results
are shown for $L=50$. (a) Temperature dependence of the leading PCA score
obtained from raw configurations. (b) Temperature dependence of the
one-dimensional CAE latent coordinate obtained from raw configurations.}
\label{fig:rawfail}
\end{figure*}

\subsection{Thermodynamic benchmark for the transition}
Fig.~\ref{fig:thermo} shows the temperature dependence of the scalar nematic order
parameter and the probability distribution $P(S)$ near coexistence. Here, $\langle S\rangle$ denotes the Monte Carlo estimate of the
ensemble average of $S$ at fixed temperature. For all
three lattice sizes, $\langle S\rangle$ decreases sharply over a narrow
temperature interval. At the same time, $P(S)$ develops a bimodal structure
near the transition, with one peak associated with the weakly ordered
nematic phase and the other with the nearly isotropic phase.

The probability distributions shown near coexistence were constructed from
configurations in a finite-size temperature window
$
|T-T_0(L)|\leq \Delta T_{\rm win}$,
with $\Delta T_{\rm win}=0.0006$. This window was chosen to improve the
statistics of the probability distributions while remaining restricted
to the narrow temperature range over which the order parameter changes
rapidly. The center temperatures were selected using guidance from the
high-resolution finite-size analysis of Ref.~\cite{Xue2024} together
with the present Monte Carlo data, yielding $T_0(50)=1.122700$, $T_0(60)=1.122659$,
and $T_0(80)=1.122496$.
The same temperature windows are used below for the thermodynamic, PCA, and CAE distributions, enabling a direct comparison of conventional and ML observables over the same transition region.

For temperature-dependent averages, error bars denote the standard error
of the mean over configurations at the same temperature. This convention is
used for the thermodynamic order parameter, PCA score, CAE latent
coordinate, and CNN-predicted order parameter. For the 
near-coexistence probability distributions, curves show the bootstrap mean
of Gaussian-smoothed histograms, and shaded bands denote
one-standard-deviation bootstrap uncertainty bands obtained from 
resamplings of the selected configurations.

These features provide the reference behavior that the ML
observables should reproduce. In particular, a useful ML diagnostic should
not merely vary smoothly with temperature; it should capture the rapid
change across the transition and, near coexistence, reflect the two-phase
structure through a bimodal distribution. 
The sharpening of the bimodal structure with increasing $L$ is consistent
with the finite-size behavior expected at a first-order transition
\cite{binder1984finite,challa1986finite} and with recent
high-precision results for the three-dimensional LL model~\cite{Xue2024}.

\subsection{Failure of unsupervised learning on raw configurations}
We next ask whether unsupervised learning can identify the transition
directly from the raw microscopic molecular field. Fig.~\ref{fig:rawfail} shows the results
obtained by applying PCA and the CAE to raw configurations. Representative results are shown for $L=50$; the same qualitative behavior was observed for $L=60$ and $L=80$.
Neither method produces a transition-sensitive variable. The leading PCA score fluctuates
without a systematic temperature dependence, and the one-dimensional CAE
latent variable remains similarly unstructured across the sampled
temperature range.

This negative result is  central to the analysis that follows. PCA and the CAE are
very different algorithms: PCA is a linear variance-based method, whereas
the CAE is a nonlinear reconstruction-based neural network. Their common
failure on the raw representation therefore indicates that the problem is
not simply insufficient model complexity. 
Instead, the problem is that the raw molecular-orientation representation does not reflect the equivalence of nematic states.

The origin of this failure lies in the symmetry of the LL model. Nematic
order is apolar and continuously degenerate. Configurations related by
global rotations, or by reversal of molecular axes, can represent the same
macroscopic nematic state while appearing numerically distinct as raw
three-component molecular fields. As a result, distances and variances in the
raw input space are dominated by irrelevant orientation differences rather
than by the scalar degree of nematic ordering. Unsupervised methods
therefore do not automatically organize the configurations according to the 
relevant collective variable.

\subsection{Symmetry-compatible bond-$P_2$ fields recover the transition}

The outcome changes qualitatively when the same Monte Carlo configurations
are represented by the rotationally invariant bond-$P_2$ field. This
representation removes the irrelevant global orientation while retaining
the local nematic alignment that appears directly in the LL interaction.
It also respects the apolar symmetry of the molecular axes and does not use
the global scalar order parameter as an input.

\begin{figure*}[htb]
\centering
 \includegraphics[width=\textwidth]{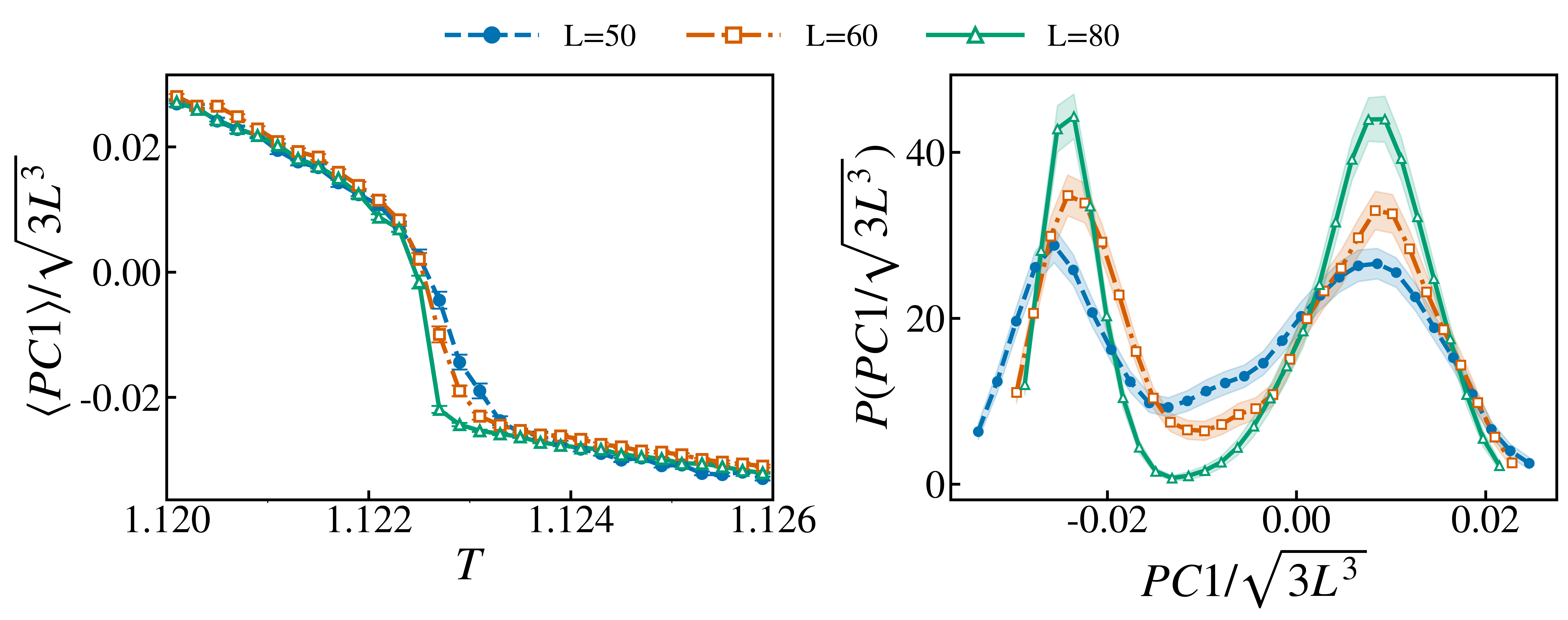}
\caption{
PCA results obtained from the rotationally invariant bond-$P_2$
representation. (a) Temperature dependence of the leading PCA score, scaled
as $PC1/\sqrt{3L^3}$ for comparison across lattice sizes. (b)
Near-coexistence probability distribution
$P(PC1/\sqrt{3L^3})$ constructed using the same finite-size temperature
windows as in Fig.~\ref{fig:thermo}.}
\label{fig:pca}
\end{figure*}

Fig.~\ref{fig:pca} shows the PCA results obtained from the bond-$P_2$
representation. The first principal component now varies sharply in the
same temperature region where the scalar order parameter changes rapidly.
Near coexistence, the distribution of the leading PCA score becomes
bimodal. Thus, after the input is made symmetry-compatible, the leading
linear variance direction behaves as an unsupervised proxy for the
transition.

\begin{figure*}[htb]
\centering
 \includegraphics[width=\textwidth]{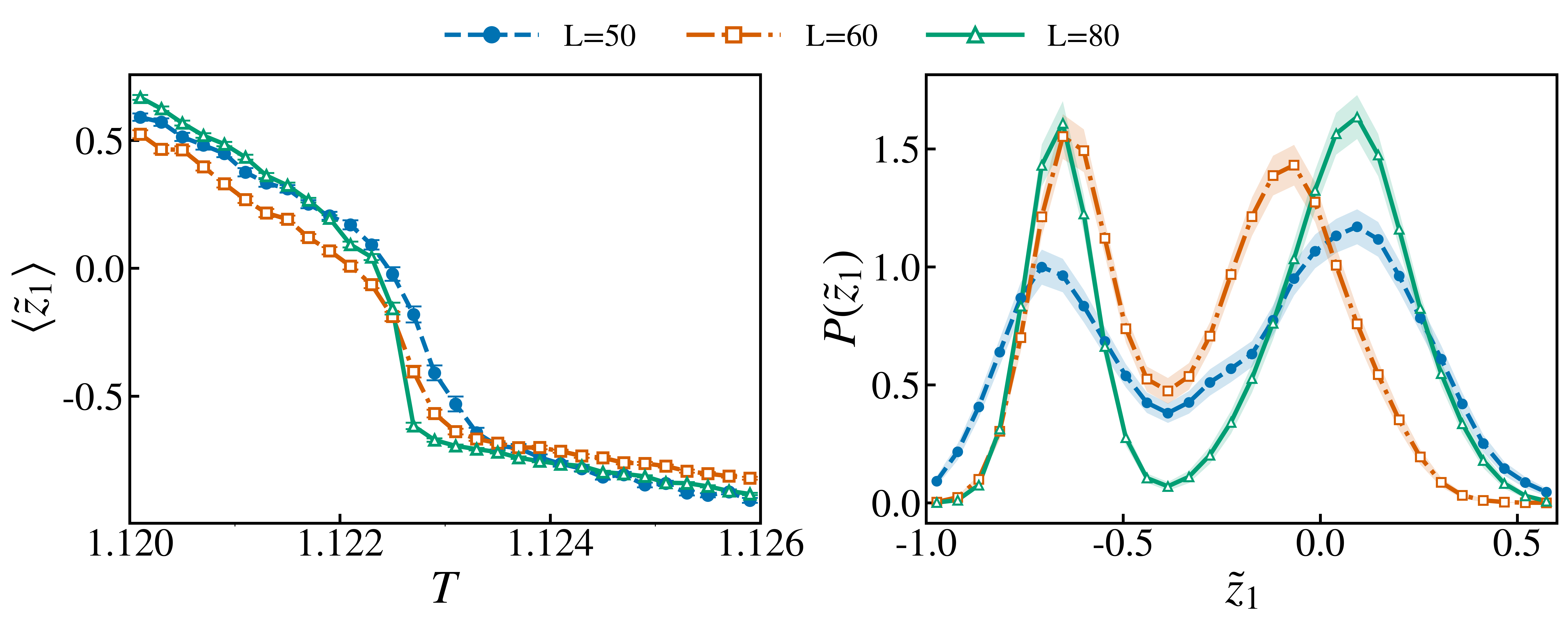}
\caption{
CAE results obtained from the rotationally invariant bond-$P_2$
representation. (a) Temperature dependence of the standardized latent
coordinate $\tilde{Z_1}=(z-\langle z\rangle_L)/\sigma_L(z)$ for $L=50$, $60$, and
$80$. (b) Near-coexistence probability distribution $P(\tilde{Z_1})$ constructed
using the same finite-size temperature windows as in Fig.~\ref{fig:thermo}.}
\label{fig:cae}
\end{figure*}

The CAE gives the same qualitative conclusion. As shown in Fig.~\ref{fig:cae}, the
one-dimensional latent coordinate changes rapidly across the transition and
develops a bimodal distribution near coexistence. This behavior mirrors the
coexistence structure seen in the thermodynamic order-parameter distribution
in Fig.~\ref{fig:thermo}, but it should not be interpreted as proving that the CAE has
learned the scalar order parameter itself. Because the CAE is trained only
to reconstruct the input field, its latent coordinate may encode a
combination of local nematic alignment, spatial correlations, and domain
structure. The important point is instead that, once the input is made
symmetry-compatible, the learned coordinate becomes a transition-sensitive
unsupervised diagnostic without using phase labels or scalar
order-parameter values.

Both a linear method and a nonlinear neural autoencoder fail on the raw representation but succeed on the bond-$P_2$ representation. The decisive factor is therefore not the
expressive power of the model alone. Rather, the transition becomes detectable by unsupervised learning only when the input representation is compatible
with the symmetry and collective variables of the nematic phase.

To our knowledge, this bond-$P_2$
representation has not previously been used in a systematic ML comparison
of raw and symmetry-compatible inputs for the microscopic three-dimensional
LL transition. Although
demonstrated here for the three-dimensional LL model, the same
representation principle is expected to be applicable to other
orientationally ordered systems with apolar continuous symmetry,
including lattice nematic models and off-lattice models of rod-like or
ellipsoidal particles.

\subsection{Supervised CNN regression from raw configurations}

\begin{figure*}[htb]
\centering
 \includegraphics[width=\textwidth]{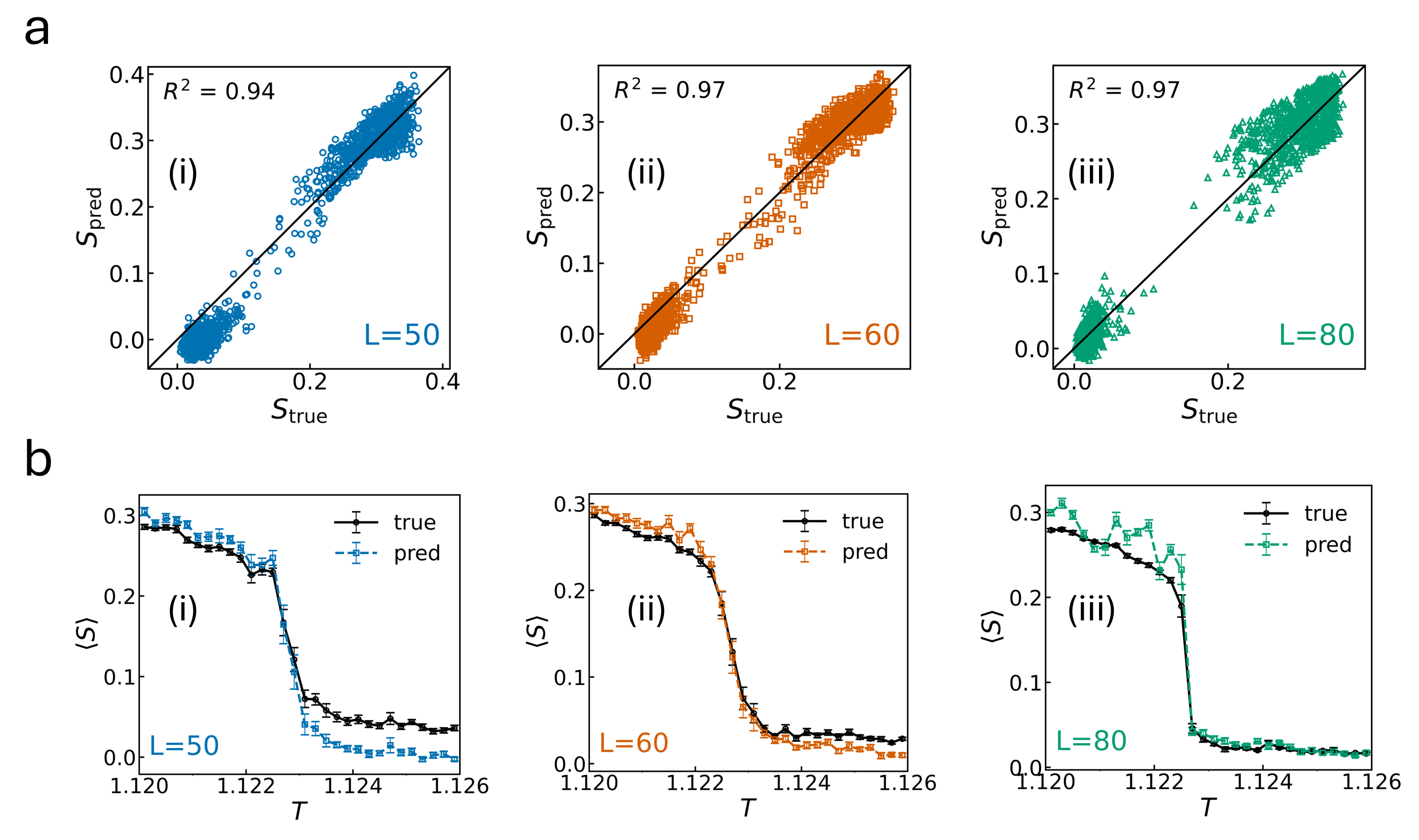}
\caption{Supervised CNN regression using raw molecular configurations as input.
(a) Predicted versus true scalar nematic order parameter for
(i) $L=50$, (ii) $L=60$, and (iii) $L=80$ on the held-out test set.
The diagonal line indicates perfect prediction. (b) Temperature dependence
of the true and predicted test-set averages of the scalar order parameter
for the same three lattice sizes.}
\label{fig:cnn}
\end{figure*}

The supervised CNN provides a complementary comparison. Unlike PCA and the
CAE, the CNN is trained with the scalar nematic order parameter as a target.
It is therefore not required to discover the relevant collective variable
from the input geometry alone. Fig.~\ref{fig:cnn} compares the predicted and true
values of $S$ for all three lattice sizes and shows the corresponding
temperature dependence of the averaged predictions.

The CNN predicts the scalar order parameter accurately from raw
configurations, with coefficients of determination $R^2(L=50)=0.94$, $R^2(L=60)=0.97$, and $R^2(L=80)=0.97$.
The averaged predictions follow the Monte Carlo order parameter closely,
including the sharp drop in the transition region.

This result should be interpreted differently from the unsupervised PCA
and CAE results. The success of the CNN does not imply that the raw
configuration space is organized by nematic order. Instead, the
target labels supply the missing information: during supervised
training, the network is driven to develop internal features correlated
with the rotationally invariant scalar order parameter. In this sense, the
CNN circumvents  the representation problem through supervision, whereas PCA
and the CAE expose the representation problem directly.

\subsection{Representation, model complexity, and supervision}

Taken together, the results separate three effects that are often
conflated in ML studies of phase transitions. First, representation is
decisive for unsupervised learning in the 3D LL model. Raw molecular
configurations do not provide a geometry in which rotationally equivalent
nematic states are close to one another, and both PCA and the CAE fail in
that representation. Second, model complexity by itself is not sufficient:
the nonlinear CAE does not recover the transition from raw configurations
any more reliably than PCA. Third, supervision changes the problem. A CNN
trained with order-parameter labels can learn from raw configurations, but
its success relies on information that is unavailable in an unsupervised
setting.

The central conclusion is , therefore , not simply that ML can
detect the isotropic--nematic transition. Rather, the LL model shows that
the visibility of the transition to ML depends on how the learning problem
is formulated. For unsupervised methods, the input representation must respect
the symmetries of the phase. For supervised methods, labels can
guide the network toward symmetry-relevant features, but at the cost of
requiring prior knowledge of the target observable.

This makes the three-dimensional LL model a useful benchmark for
ML studies of systems with continuous orientational
symmetry. The ordered phase is not characterized by an absolute molecular
orientation, but by rotationally invariant nematic alignment. 
The bond-$P_2$ representation makes this structure explicit, allowing both linear and nonlinear unsupervised methods to recover
the transition signatures observed in conventional thermodynamic quantities.

\section{Conclusions}
\label{sec:conclsuion}
We have studied the weak first-order isotropic--nematic transition of the
three-dimensional Lebwohl--Lasher model using Monte Carlo simulations and
three complementary ML methods. Conventional thermodynamic
observables show the expected sharp change in the scalar nematic order
parameter and bimodal coexistence distributions. These thermodynamic
features provide the benchmark against which the ML observables were
compared.

The main result is that unsupervised learning succeeds or fails depending
on the symmetry compatibility of the input representation. PCA and the CAE
do not isolate the transition when applied to raw molecular configurations,
because rotationally equivalent nematic states can appear far apart in the
raw representation. When the same configurations are transformed into a
rotationally invariant bond-$P_2$ field, both methods recover sharp
transition signatures and bimodal distributions near coexistence. The
agreement between PCA and the CAE shows that the effect is
representation-driven rather than architecture-specific.

The supervised CNN provides a complementary result. Trained directly on raw
configurations, it accurately predicts the scalar nematic order parameter
for all three lattice sizes,  with $R^2 \simeq 0.94$ or higher. This demonstrates that
supervision can guide a neural network toward symmetry-relevant features,
even when the input representation itself is not explicitly invariant.

These results show that ML can serve as a reliable probe of the LL isotropic--nematic transition, but also that unsupervised ML requires careful attention to input representation. More broadly, the choice of input representation is part of the physics problem, not merely a preprocessing step.
For systems with orientational or continuous symmetries, unsupervised ML works reliably
only when the input representation removes symmetry-equivalent degrees of
freedom and retains the local invariants associated with the ordered phase.

\section*{Acknowledgments}
The authors thank Dilina Perera for a careful reading of the manuscript and for valuable comments.
\section*{Data Availability}
The data that support the findings of this study are available from the authors upon reasonable request.

\bibliographystyle{unsrtnat}
\bibliography{ll_ml_refs}

@article{metropolis1953,
  author    = {N. Metropolis and A. W. Rosenbluth and M. N. Rosenbluth and A. H. Teller and E. Teller},
  title     = {Equation of state calculations by fast computing machines},
  journal   = {The Journal of Chemical Physics},
  volume    = {21},
  year      = {1953},
  pages     = {1087--1092},
  doi       = {10.1063/1.1699114},
  url =       {https://doi.org/10.1063/1.1699114}
}

@article{LebwohlLasher1972,
  title = {Nematic-Liquid-Crystal Order---A Monte Carlo Calculation},
  author = {Lebwohl, P. A. and Lasher, G.},
  journal = {Physical  Review A},
  volume = {6},
  pages = {426--429},
  numpages = {0},
  year = {1972},
  publisher = {American Physical Society},
  doi = {10.1103/PhysRevA.6.426},
  url = {https://link.aps.org/doi/10.1103/PhysRevA.6.426}
}

@article{binder1984finite,
  author  = {Binder, K. and Landau, D. P.},
  title   = {Finite-size scaling at first-order phase transitions},
  journal = {Physical Review B},
  volume  = {30},
  pages   = {1477--1485},
  year    = {1984},
  doi     = {10.1103/PhysRevB.30.1477},
  url     = {https://doi.org/10.1103/PhysRevB.30.1477}
}

@article{challa1986finite,
  author  = {Challa, Murty S. S. and Landau, D. P. and Binder, K.},
  title   = {Finite-size effects at temperature-driven first-order transitions},
  journal = {Physical Review B},
  volume  = {34},
  pages   = {1841--1852},
  year    = {1986},
  doi     = {10.1103/PhysRevB.34.1841},
  url    =  {https://doi.org/10.1103/PhysRevB.34.1841}
}

@inproceedings{Masci2011,
  author    = {Masci, Jonathan and Meier, Ueli and Cire{\c{s}}an, Dan C. and Schmidhuber, J{\"u}rgen},
  title     = {Stacked Convolutional Auto-Encoders for Hierarchical Feature Extraction},
  booktitle = {Artificial Neural Networks and Machine Learning -- ICANN 2011},
  series    = {Lecture Notes in Computer Science},
  volume    = {6791},
  pages     = {52--59},
  publisher = {Springer},
  year      = {2011},
  doi       = {10.1007/978-3-642-21735-7_7},
  url       = {https://doi.org/10.1007/978-3-642-21735-7_7}
}

@inproceedings{Kingma2015,
  author    = {Kingma, Diederik P. and Ba, Jimmy},
  title     = {Adam: A Method for Stochastic Optimization},
  booktitle = {International Conference on Learning Representations},
  year      = {2015},
  url       = {https://arxiv.org/abs/1412.6980}
}

@article{Wang2016,
  author  = {Wang, Lei},
  title   = {Discovering phase transitions with unsupervised learning},
  journal = {Physical Review B},
  volume  = {94},
  pages   = {195105},
  year    = {2016},
  doi     = {10.1103/PhysRevB.94.195105},
  url     = {https://link.aps.org/doi/10.1103/PhysRevB.94.195105}
}

@article{Jolliffe2016,
  author  = {Jolliffe, Ian T. and Cadima, Jorge},
  title   = {Principal component analysis: a review and recent developments},
  journal = {Philosophical Transactions of the Royal Society A},
  volume  = {374},
  pages   = {20150202},
  year    = {2016},
  doi     = {10.1098/rsta.2015.0202},
  url     = {https://doi.org/10.1098/rsta.2015.0202}
}

@article{CarrasquillaMelko2017,
  author  = {Carrasquilla, Juan and Melko, Roger G.},
  title   = {Machine learning phases of matter},
  journal = {Nature Physics},
  volume  = {13},
  pages   = {431--434},
  year    = {2017},
  doi     = {10.1038/nphys4035},
  url     = {https://www.nature.com/articles/nphys4035}
}

@article{vanNieuwenburg2017,
  author  = {van Nieuwenburg, Evert P. L. and Liu, Ye-Hua and Huber, Sebastian D.},
  title   = {Learning phase transitions by confusion},
  journal = {Nature Physics},
  volume  = {13},
  pages   = {435--439},
  year    = {2017},
  doi     = {10.1038/nphys4037},
  url     = {https://www.nature.com/articles/nphys4037}
}

@article{Hu2017,
  author  = {Hu, Wenjian and Singh, Rajiv R. P. and Scalettar, Richard T.},
  title   = {Discovering phases, phase transitions, and crossovers through unsupervised machine learning: A critical examination},
  journal = {Physical Review E},
  volume  = {95},
  pages   = {062122},
  year    = {2017},
  doi     = {10.1103/PhysRevE.95.062122},
  url     = {https://link.aps.org/doi/10.1103/PhysRevE.95.062122}
}

@article{Wetzel2017,
  author  = {Wetzel, Sebastian J.},
  title   = {Unsupervised learning of phase transitions: From principal component analysis to variational autoencoders},
  journal = {Physical Review E},
  volume  = {96},
  pages   = {022140},
  year    = {2017},
  doi     = {10.1103/PhysRevE.96.022140},
  url     = {https://link.aps.org/doi/10.1103/PhysRevE.96.022140}
}

@article{Broecker2017,
  author  = {Broecker, Peter and Carrasquilla, Juan and Melko, Roger G. and Trebst, Simon},
  title   = {Machine learning quantum phases of matter beyond the fermion sign problem},
  journal = {Scientific Reports},
  volume  = {7},
  pages   = {8823},
  year    = {2017},
  doi     = {10.1038/s41598-017-09098-0},
  url     = {https://www.nature.com/articles/s41598-017-09098-0}
}

@article{Jadrich2018a,
  author  = {Jadrich, R. B. and Lindquist, B. A. and Truskett, T. M.},
  title   = {Unsupervised machine learning for detection of phase transitions in off-lattice systems. I. Foundations},
  journal = {The Journal of Chemical Physics},
  volume  = {149},
  pages   = {194109},
  year    = {2018},
  doi     = {10.1063/1.5049849},
  url     = {https://doi.org/10.1063/1.5049849}
}

@article{Jadrich2018b,
  author  = {Jadrich, R. B. and Lindquist, B. A. and Pi\~neros, W. D. and Banerjee, D. and Truskett, T. M.},
  title   = {Unsupervised machine learning for detection of phase transitions in off-lattice systems. II. Applications},
  journal = {The Journal of Chemical Physics},
  volume  = {149},
  pages   = {194110},
  year    = {2018},
  doi     = {10.1063/1.5049850},
  url     = {https://doi.org/10.1063/1.5049850}
}

@article{Sigaki2019PRE,
  author = {Sigaki, Higor Y. D. and de Souza, R. F. and de Souza, R. T. and Zola, Rafael S. and Ribeiro, Haroldo V.},
  title = {Estimating physical properties from liquid crystal textures via machine learning and complexity-entropy methods},
  journal = {Physical Review E},
  volume = {99},
  pages = {013311},
  year = {2019},
  doi = {10.1103/PhysRevE.99.013311},
  url ={https://journals.aps.org/pre/abstract/10.1103/PhysRevE.99.013311}
}

@article{Shiina2020,
  author  = {Shiina, Kenta and Mori, Hiroyuki and Okabe, Yutaka and Lee, Hwee Kuan},
  title   = {Machine-Learning Studies on Spin Models},
  journal = {Scientific Reports},
  volume  = {10},
  pages   = {2177},
  year    = {2020},
  doi     = {10.1038/s41598-020-58263-5},
  url     = {https://www.nature.com/articles/s41598-020-58263-5}
}

@article{Sigaki2020,
  author  = {Sigaki, Higor Y. D. and Lenzi, Ervin K. and Zola, Rafael S. and Perc, Matja\v{z} and Ribeiro, Haroldo V.},
  title   = {Learning physical properties of liquid crystals with deep convolutional neural networks},
  journal = {Scientific Reports},
  volume  = {10},
  pages   = {7664},
  year    = {2020},
  doi     = {10.1038/s41598-020-63662-9},
  url     = {https://www.nature.com/articles/s41598-020-63662-9}
}

@book{landau2021guide,
  author    = {Landau, David P. and Binder, Kurt},
  title     = {A Guide to Monte Carlo Simulations in Statistical Physics},
  edition   = {5},
  publisher = {Cambridge University Press},
  address   = {Cambridge},
  year      = {2021}
}

@article{NgYang2023AutoencodersPhaseTransitions,
  title   = {Unsupervised learning of phase transitions via modified anomaly detection with autoencoders},
  author  = {Ng, Kwai-Kong and Yang, Min-Fong},
  journal = {Physical Review B},
  volume  = {108},
  pages   = {214428},
  year    = {2023},
  doi     = {10.1103/PhysRevB.108.214428},
  url     = {https://doi.org/10.1103/PhysRevB.108.214428},
}

@article{Xue2024,
  author  = {Xue, Aojie and Xu, Jiahao and Landau, D. P. and Binder, K.},
  title   = {Test of universality at first order phase transitions: The Lebwohl--Lasher model},
  journal = {The Journal of Chemical Physics},
  volume  = {161},
  pages   = {134107},
  year    = {2024},
  doi     = {10.1063/5.0221215},
  url     = {https://doi.org/10.1063/5.0221215}
}

@article{Piven2024,
  author  = {Piven, Anastasiia and Darmoroz, Darina and Skorb, Ekaterina and Orlova, Tetiana},
  title   = {Machine learning methods for liquid crystal research: phases, textures, defects and physical properties},
  journal = {Soft Matter},
  year    = {2024},
  volume  = {20},
  pages   = {1380-1391},
  doi     = {10.1039/d3sm01634j},
  url     = {https://doi.org/10.1039/d3sm01634j},
}

@article{Beyerle2025,
  author  = {Beyerle, Eric R. and Tiwary, Pratyush},
  title   = {Inferring the Isotropic--Nematic Phase Transition with Generative Machine Learning},
  journal = {Physical Review Letters},
  volume  = {135},
  pages   = {068102},
  year    = {2025},
  doi     = {10.1103/1wdj-ym3s},
  url     = {https://link.aps.org/doi/10.1103/1wdj-ym3s}
}

@article{maninder2026,
  title = {Can machine learning truly decode phase transitions? A deep dive into the Ising model with competing interactions},
  author = {Kaur, Maninder and Li, Ying Wai and Perera, Dilina and Landau, David P.},
  journal = {Phys. Rev. E},
  volume = {114},
  pages = {014104},
  numpages = {11},
  year = {2026},
  publisher = {American Physical Society},
  doi = {10.1103/49zd-8dgw},
  url = {https://link.aps.org/doi/10.1103/49zd-8dgw}
}

\end{document}